# Demand allocation with latency cost functions

Alessandro Agnetis    Enrico Grande    Andrea Pacifici

October 30, 2018

**Abstract**

We address the exact resolution of a MINLP model where resources can be activated in order to satisfy a demand (a partitioning constraint) while minimizing total cost. Cost functions are convex latency functions plus a fixed activation cost. A branch and bound algorithm is devised, featuring three important characteristics. First, the lower bound (therefore each subproblem) can be computed in $O(n \log n)$. Second, to break symmetries resulting in improved efficiency, the branching scheme is $n$-ary (instead of the "classical" binary). Third, a very affective heuristic is used to compute a good upper bound at the root node of the enumeration tree. All three features lead to a successful comparison against CPLEX MIPQ, which is the fastest among several commercial and open-source solvers: computational results showing this fact are provided.

## 1 Introduction

Congestion phenomena arise in many real world problems. It is therefore a key issue in modeling techniques to address those features. Latency functions (see [7]) are a useful tool to model nonlinear increase of the cost of a resource with respect to its usage: as an example think of traffic jams. Traffic problems are widely studied applying the concept of equilibrium (see [5]). Studies about behaviors of (possibly) congested systems where users act autonomously, i.e. selfishly, have addressed the problem of quantifying the loss of efficiency compared to a centrally optimized situation (see [8]). There's also been efforts in the direction of pricing edges of networks in order to influence users behaviors (see [4]).

Our model addresses optimization problems where there is (i) a single commodity demand that has to be satisfied, (ii) a set of potential resources, (iii) a fixed activation cost for each resource and (iv) the congestion heavily influences the cost of a resource. We aim to a *socially desirable* solution, that is, a global optimum.

We denote by $Q$ the set of the $q$ available resources. The cost $\gamma_i(x_i)$ of using resource $i \in Q$ at level $x_i \in [0,1]$ is given by:

$$\gamma_i(x_i) = \begin{cases} 0 & \text{if } x_i = 0 \\ c_i + x_i f_i(x_i) & \text{if } 0 < x_i \leq 1 \end{cases} \quad (1)$$

where $f_i(\cdot)$—the latency function—satisfies the following:

**Assumption 1** $f_i(\cdot)$ *is semi-convex with* $f_i(0) = 0$ *and its derivative* $\dfrac{df_i(\xi)}{d\xi} > 0$ *for* $\xi > 0$.

Note that, for any $i \in Q$, the function $c_i + x_i f(x_i)$ is not convex in the domain $x_i \geq 0$, although it is for $x_i > 0$. Unless differently specified, we assume troughout the paper that latency functions $f_i(\cdot)$ satisfy Assumption 1

The model is the following:

$$z^* = \min \quad \sum_{i \in Q} c_i y_i + x_i f_i(x_i) \tag{2}$$

$$\text{s.t.} \quad x_i \leq y_i, \text{ for all } i \in Q \tag{3}$$

$$\sum_{i \in Q} x_i = 1 \tag{4}$$

$$x \in \mathbb{R}_+^q \tag{5}$$

$$y \in \{0,1\}^q \tag{6}$$

where $x_i$ indicates the fraction of demand allocated to resource $i$ (level of usage of resource $i$) and $y_i$ is a binary variable indicating wether resource $i$ is active ($x_i > 0 \Rightarrow y_i = 1$) or not ($x_i = 0 \Rightarrow y_i = 0$). The above formulation is a mixed integer nonlinear program and, in particular, all the constraints are linear with the $y$ variables restricted to be integer. We will often refer to (2–6) as $P$.

## 2 Complexity issues

In this section we prove that $P$ is, in general, a $\mathcal{NP}$-hard problem. In fact even when $f(\cdot)$ is a homogeneous linear function $P$ contains PARTITION, as we prove in Section 2.2. Nevertheless, as discussed respectively in Section 2.1.1, 2.1.2 and 2.1.3, the problem is easy when the cost functions are identical, or there are null fixed costs (i.e., $c_i = 0$ for all $i \in Q$) or when $f(\cdot)$ is constant.

### 2.1 Easy cases

#### 2.1.1 Identical cost functions

We consider the case when all cost functions $\gamma_i(x_i)$ are of the same type, that is, for all $i \in Q$, $\gamma_i(x_i) = c + x_i f(x_i) = \gamma(x_i)$. To the best of our knowledge, these results, though straightforward, are not present in the literature. In this case, the problem reduces to finding the optimal number $k \leq q$ of active resources.

**Proposition 1** *If the resources are all of the same type with cost functions $\gamma_i(\cdot) = \gamma(\cdot)$, for all $i \in Q$, and an optimal solution for $P$ consists of using $k$ resources, then there is one solution where equal fraction of demand is allocated to the $k$ resources.*

**Proof.** Based on the convexity of $x_i f_i(x_i)$, for any $k$-uple of nonnegative numbers $x_1, \ldots, x_k$, with $\sum_{i=1}^{k} x_i = 1$, we have:

$$k \cdot \gamma\left(\frac{1}{k}\right) \leq \sum_{i=1}^{k} \gamma(x_i). \tag{7}$$

∎

The next natural question we need to ask is "What is an optimal number of such resources (i.e., the best value for k)?". It is easy to observe that with zero setup costs ($c = 0$) the cheapest solution consists of using the maximum number (i.e., $q$) of resources.

A solution that uses $k + 1$ resources costs no more than a solution with $k$ resources if and only if the following is true:

$$c + \frac{1}{k+1} f\left(\frac{1}{k+1}\right) \leq \frac{1}{k} f\left(\frac{1}{k}\right). \qquad (8)$$

If $c = 0$, by definition of $f(\cdot)$, the last inequality is always valid. Therefore, it is cost-effective to use another resource if the additional setup cost does not exceed the gain in the variable costs.

The cost of using $k$ resources is

$$F(k) = k\left(\frac{1}{k}\right) f\left(\frac{1}{k}\right) + kc = f\left(\frac{1}{k}\right) + kc. \qquad (9)$$

If we define $g(\theta) = \theta f(\theta)$, which is convex by definition of $f(\cdot)$, we have that

$$\frac{\partial^2 F}{\partial k^2} = \frac{\partial^2}{\partial k^2}\left(kg\left(\frac{1}{k}\right)\right) = \left(\frac{1}{k^3}\right) \frac{\partial^2 g}{\partial k^2}\bigg|_{\frac{1}{k}} \geq 0 \qquad (10)$$

Hence, there must be $1 \leq k^* \leq q$ such that

$$F(1) \geq F(2) \geq \ldots \geq F(k^*) \text{ and } F(k^*) \leq F(k^* + 1) \leq \ldots \leq F(q). \qquad (11)$$

Therefore, since a binary search can be used to efficiently find the $k^*$, the following proposition holds.

**Proposition 2** *When the $q$ resources are all of the same type with the cost functions $\gamma_i(\cdot) = \gamma(\cdot)$, for all $i \in Q$, the problem is solvable in $O(C \log(q))$ time, where $C$ is the maximum computational effort for calculating $\gamma(\frac{1}{k})$.* ■

### 2.1.2 Null fixed costs

When the fixed costs are all null, then the objective function reduces to a sum of convex functions $x_i f_i(x_i)$, which is also convex. Since there is no need of binary variables to deal with fixed costs, we are left to the problem of minimizing a convex function over a polyhedron and this can be achieved via Karush-Kuhn-Tucker conditions. The optimal solution of this problem is also referred to as a Nash equilibrium (see [2]), because the partial derivatives of the cost functions are all set to the same value.

### 2.1.3 Constant latency functions

The objective becomes a linear function when the $f_i(x_i) = f_i$ are given constans. In this case an optimal solution is simply $x_{i^*}^* = y_{i^*}^* = 1$ where $i^* = \arg\min\{c_i + f_i, \ i \in Q\}$ and $x_i^* = y_i^* = 0$ for all $i \neq i^*$.

## 2.2 NP-hardness of $P$

From the discussion in Section 2.1, the minimal case for which the problem is possibly $\mathcal{NP}$-hard is when there are $q$ resources $f_i(x) = b_i x$.

**Proposition 3** *Problem 2–6 is $\mathcal{NP}$-hard even when the latency functions $f_i(\cdot)$ are linear.*

**Proof.** We use a reduction from the well known binary $\mathcal{NP}$-hard problem

> PARTITION: *given $q$ nonnegative integers $w_1, w_2, \ldots, w_q$, with $\sum_{i=1}^{q} w_i = W$, find, if it exists, a subset $S \subseteq \{1, \ldots, q\}$ such that $\sum_{i \in S} w_i = \frac{W}{2}$.*

First, note that when we select a set of resources, the following convex subproblem may be easily solved by applying the Karush-Kuhn-Tucker (KKT) optimality conditions. Let $S \subseteq Q$ be the set of selected resources. Then Problem 2–6 restricts to

$$z(S) = \min \left\{ \sum_{i \in S} f_i + b_i x_i^2 : \sum_{i \in S} x_i = 1; x_i \in \mathbb{R}_+^n \right\}. \tag{12}$$

Restricted problem 12 is a convex optimization problem. Let us construct the following Lagrangean function (for simplicity the constant term $\sum_{i \in S} f_i$ has been omitted in the objective function below):

$$L(x, \mu, \lambda) = \sum_{i \in S} b_i x_i^2 + \lambda \left(1 - \sum_{i \in S} x_i \right) + \sum_{i \in S} \mu_i x_i$$

The KKT conditions are (the symbol * indicates the optimal solution)

$$\nabla_x L(x^*, \lambda^*, \mu^*) = 0_q$$
$$\sum_{i \in S} x_i = 1$$
$$\mu^{*T} x^* = 0$$
$$x^* \geq 0_q$$
$$\mu^* \geq 0_q$$

We then obtain

$$\lambda^* = \frac{1}{\sum_{i \in S} \frac{1}{2b_i}}$$

and choosing $x_i^* \neq 0 \Rightarrow \mu_i = 0 \quad \forall i \in S$, we have

$$x_i^* = \frac{\lambda^*}{2b_i} = \frac{1}{2b_i \sum_{j \in S} \frac{1}{2b_j}}$$

These values satisfy the KKT conditions and are therefore a global optimum for Problem 12.

Consider now an instance of Problem 2–6 with $f_i = w_i$ and $b_i = \frac{1}{w_i}\left(\frac{W}{4}\right)$, where the $w_i$'s and $W$ are those of the PARTITION instance. Considering the above Restricted formulation and given the set of active resources $S$, the optimal solution value can be rewritten as

$$z(S) = \min \left\{ \sum_{i \in S} w_i + \frac{\frac{W^2}{4}}{\sum_{i \in S} w_i} \right\}$$

Observe now that the function $\xi + \frac{a}{\xi}$ has a minimum at $\xi^* = \sqrt{a}$, therefore $z(S)$ is minimized when $\sum_{i \in S} w_i = \sqrt{\frac{W^2}{4}} = \frac{W}{2}$. Hence Problem (2)–(6) can be rewritten as

$$\min_{S \subseteq \{1,\ldots,q\}} z(S) \tag{13}$$

which is equivalent to finding $S \subseteq \{1,\ldots,q\}$ such that $\sum_{i \in S} w_i = \frac{W}{2}$. This is, in turn, equivalent to PARTITION. This completes the proof. ∎

## 3 Dual Bounds

In this Section we will study the Lagrangean relaxation of the problem, its Continuous relaxation and the relation between them, in order to efficiently compute a dual bound. For the development to follow, we first address the case when the set $S \subseteq Q$ of selected resources (i.e. active facilities) is given. In this case, the fixed costs are known and variables $y_i$, $i \in Q$, vanish from (2–6). The resulting problem can be written as:

$$z(S) = \sum_{i \in S} c_i + \min \left\{ \sum_{i \in S} x_i f_i(x_i) \ : \ \sum_{i \in S} x_i = 1;\ x_i \in \mathbb{R}_+,\ i \in S \right\} \tag{14}$$

and, since its objective function is convex, it may solved by applying Karush-Kuhn-Tucker optimality (KKT) conditions.

Omitting the constant term $\sum_{i \in S} c_i$, the Lagrangean function

$$L(x, \mu, \lambda) = \sum_{i \in S} x_i f_i(x_i) + \lambda \left( 1 - \sum_{i \in S} x_i \right) - \sum_{i \in S} \mu_i x_i$$

The KKT conditions, for the triple $(x^*, \mu^*, \lambda^*)$, are

$$\frac{\partial L(x^*, \lambda^*, \mu^*)}{\partial x_i} = f_i(x_i^*) + x_i^* f_i'(x_i^*) - \lambda^* - \mu_i^* = 0 \quad \text{for all } i \in S$$

$$\sum_{i \in S} x_i^* = 1$$

$$\mu_i^* x_i^* = 0 \quad \text{for all } i \in S$$

$$x_i^* \geq 0 \quad \text{for all } i \in S$$

$$\mu_i^* \geq 0 \quad \text{for all } i \in S$$

From $x_i^* > 0$ we have that $\mu_i^* = 0$ for all $i \in S$. Then,

$$x_i^* = g_i^{-1}(\lambda^*) \tag{15}$$

where $g_i(\zeta) = f_i(\zeta) + \zeta f_i'(\zeta)$—the derivative of $\zeta f_i(\zeta)$—is a monotonically increasing function[1] and $\lambda^*$ can be computed, independently of the $x_i^*$'s, by

$$\sum_{i \in S} g_i^{-1}(\lambda^*) = 1. \tag{16}$$

---
[1] Here we neglect the case constant $f_i$.

Depending on the functions $f_i(\cdot)$, the value of $\lambda^*$ and $x_i^*$ can be computed in closed form or via numerical methods. For instance, if $f_i(\zeta) = b_i \zeta^p$, we have that:

$$\lambda^* = \left( \frac{1}{\sum_{i \in S} \frac{1}{\sqrt[p]{b_i(1+p)}}} \right)^p \quad \text{and} \quad x_i^* = \sqrt[p]{\frac{\lambda^*}{b_i(1+p)}}.$$

## 3.1 Lagrangean relaxation

We now use Lagrangean relaxation to obtain a lower bound on $z^*$, the optimal solution value of Problem (2–6).

Relaxing the activation constraints (3) using nonnegative Lagrangean multipliers $\kappa_i, i = 1, \ldots, |Q|$, we obtain the following problem:

$$z_{LRP}(\kappa) = \min \ \sum_{i \in Q} (c_i - \kappa_i) y_i + x_i f(x_i) + \kappa_i x_i \tag{17}$$
$$\text{s.t.} \quad \sum_{i \in Q} x_i = 1 \tag{18}$$
$$x \in \mathbb{R}_+^q \tag{19}$$
$$y \in \{0,1\}^q \tag{20}$$

Problem (17–20) is a relaxation of the original Problem for any $\kappa \geq 0_q$, and it is decomposable since optimal values for the $y$ variables are independent of the values of the $x$. The optimal solution values for the $y$ variables are the following:

$$y_i^* = \begin{cases} 1 & \text{if } c_i < \kappa_i \\ 0 & \text{if } c_i \geq \kappa_i \end{cases} \quad \text{for all } i \in Q.$$

The remaining convex program, on the $x$ variables, is:

$$z_{LRP'}(\kappa) = \min \ \sum_{i \in Q} x_i f(x_i) + \kappa_i x_i \tag{21}$$
$$\text{s.t.} \quad \sum_{i \in Q} x_i = 1 \tag{22}$$
$$x \in \mathbb{R}_+^q \tag{23}$$

and therefore $z_{LRP}(\kappa) = z_{LRP'}(\kappa) + \sum_{i \in Q} (c_i - \kappa_i) y_i^*$. Problem (21–23), is a convex program and can be optimally solved via KKT conditions. Using multiplier $\lambda \in \mathbb{R}$ for the covering constraint (22) and multipliers $\mu \in \mathbb{R}_+^q$ for nonnegativity constraints (23), we obtain the following Lagrangean function:

$$L_\kappa(\lambda, \mu) = \min \sum_{i \in Q} (x_i f_i(x_i) + \kappa_i x_i - \mu_i x_i) + \lambda \left( 1 - \sum_{i \in Q} x_i \right).$$

Thus KKT conditions are

$$\frac{\partial L_k(x^*, \lambda^*)}{\partial x_i} = g_i(x_i^*) + \kappa_i - \lambda^* - \mu_i^* = 0 \quad \text{for all } i \in Q \tag{24}$$
$$\sum_{i \in S} x_i^* = 1 \tag{25}$$
$$\mu_i^* x_i^* = 0 \quad \text{for all } i \in Q \tag{26}$$
$$x_i^* \geq 0 \quad \text{for all } i \in Q \tag{27}$$
$$\mu_i^* \geq 0 \quad \text{for all } i \in Q \tag{28}$$

Now note that if $i \in S \subseteq Q$ then $x_i^* > 0$ and $\mu_i^* = 0$; since, from Assumption 1, the first derivative is positive for poitive arguments, we have that $g_i(x_i^*) = \lambda^* - \kappa_i > 0$, resulting in

$$\lambda^* > \kappa_i, \text{ for all } i \in S. \tag{29}$$

On the other hand, considering those $i \notin S$ we have that $x_i^* = 0 \Rightarrow \mu_i^* \geq 0 \Rightarrow g_i(x_i^*) = \lambda^* + \mu_i^* - \kappa_i$; the latter implication together with the fact that $g_i(0) = 0$ lead to $\mu_i^* = \kappa_i - \lambda^* \geq 0$. This implies

$$\lambda^* \leq \kappa_i, \text{ for all } i \notin S. \tag{30}$$

Equations 29 and 30 induce a particular structure for a solution to be optimal for (21–23), which is the following:

$$\underbrace{\kappa_1 \leq \kappa_2 \leq \ldots \leq \kappa_h}_{S} < \lambda^* \leq \underbrace{\kappa_{h+1} \leq \ldots \leq \kappa_q}_{Q \setminus S}. \tag{31}$$

Furthermore from 25 and from the above discussion we obtain

$$x_i^* = g_i^{-1}(\lambda^* - \kappa_i) \tag{32}$$

and since we have $\sum_{i \in S} x_i^* = 1 = \sum_{i \in S} g_i^{-1}(\lambda^* - \kappa_i)$, the expression for $\lambda^*$ is

$$\lambda^* = G(\kappa_1, \ldots, \kappa_h). \tag{33}$$

where $G(\cdot)$ can be computed in closed form (where possible) or via numerical methods.

The above results suggest an efficient algorithm to compute the optimal solution of the relaxed problem (21–23): order the $\kappa_i$ in non-decreasing sense, compute $\lambda^*$ using 33 such that 31 is satisfied, compute the optimal variables for $i \in S$ using 32 and set $x_i^* = 0$ for all $i \notin S$. We will refer to this algorithm as to the *Ordering Algorithm*.

### 3.2 Continuous relaxation

In the continuous relaxation of out original problem we relax the integrality constraint on the $y$ variables. The problem becomes:

$$w = \min \sum_{i \in Q} c_i y_i + x_i f_i(x_i) \tag{34}$$

$$\text{s.t.} \quad x_i \leq y_i, \text{ for all } i \in Q \tag{35}$$

$$\sum_{i \in Q} x_i = 1 \tag{36}$$

$$x \in \mathbb{R}_+^q \tag{37}$$

$$0 \leq y_i \leq 1, \text{ for all } i \in Q \tag{38}$$

It is easy to see that in the optimal solution of this problem $x_i^* = y_i^*$ for all $i \in Q$ and thus we can drop constraints 35 and 38 and replace $y_i$ with $x_i$ in the objective function. The problem is now the following:

$$w = \min \sum_{i \in Q} c_i x_i + x_i f_i(x_i) \tag{39}$$

$$\text{s.t.} \quad \sum_{i \in Q} x_i = 1 \tag{40}$$

$$x \in \mathbb{R}_+^q \tag{41}$$

Note that setting $c_i = \kappa_i$ makes the Problem (39–41) *equal* to Problem (21–23), that is $w = z_{LRP'}(c)$. By convexity (the proof is an extension of a well known result in LP, see [6] and [3]) we have that $w = \max_k\{z_{LRP'}(k)\}$ and therefore we have the optimal multipliers for the optimal dual solution, that is $k^* = c$. Another consequence of this observation is that the Continuous relaxation (whose optimal values equals the optimal lagrangean dual) is solvable using the Ordering Algorithm described in the previous Section.

## 4 Primal Bound

In this Section we describe a heuristic method that exploits the theory developed in the previous sections. The heuristic starts from the primal solution value induced by the optimal dual solution point (activated resources in the dual solution) and then performs a local search to improve it.

Compute the set $S \subseteq Q$ of activated resources—and therefore $\lambda^*$—as in the Ordering Algorithm; then use Equation 15 to compute values of the activated variables ($i \in S$) and set the others to zero ($x_i = 0, i \notin S$). The current objective function value (primal) is:

$$z(S) = \sum_{i \in S} c_i + x_i f_i(x_i)$$

Now perform a local (greedy) search. Select—with a suitable criterion—an activated resource $u$ to remove from $S$ and compute $z(S \setminus \{j\})$: if this solution is better than the previous, update the solution ($S := S \setminus \{j\}$). Now select a non-activated $v$ resource to include in $S$ and compute $z(S \cup \{j\})$: if this solution is better than the previous, update the solution ($S := S \cup \{j\}$).

The choice criterion for the local search can be tailored on the particular $f_i(x_i)$. Anyhow a criterion which can be always applied is the one of choosing $i \in S$ as $\arg\max\{c_i\}$ and $i \notin S$ as $\arg\min\{c_i\}$. As a example if $f_i(x_i) = b_i x_i^p$, a criterion can be to compute the $\arg\max$ of $c_i$, of $c_i + b_i$, and of $c_i + \sqrt[p+1]{b_i}$ and to pick the $\arg\max$ of them.

## 5 Exact algorithm

In this Section we devise an exact branch-and-bound algorithm for (2–6) based on the developments of the previous sections.

### 5.1 Solution strategy

In the enumeration tree, each node $\nu$ represents a subproblem that is defined by $(i)$ a set of *active* (ON) resources $T$, that is $T = \{i : y_i = 1; i \in \{1, \ldots, q\}\} \subseteq Q$, $(ii)$ a set of inactive (OFF) resources $F$ ($F \subseteq Q, F \cap T = \emptyset, F = \{1 \leq i \leq q\}$) that must be OFF and $(iii)$ the set $Q \setminus \{T \cup F\}$ of resources that are not yet fixed to ON or OFF. If a resource is OFF, we consider unavailable for that specific subproblem. We indicate by $Q(\nu) = Q \setminus F$ the set of available resources at node (subproblem) $\nu$. The generic subproblem may be formulated as follows:

$$\min \sum_{i \in Q(\nu) \setminus T} (c_i y_i + x_i f_i(x_i)) + \sum_{i \in T} (c_i + x_i f_i(x_i))$$
$$\text{s.t.} \quad x_i \leq y_i \text{ for all } i \in Q(\nu) \setminus T$$
$$\sum_{i \in Q(\nu)} x_i = 1$$
$$x_i \geq 0 \text{ for all } i \in Q(\nu)$$
$$y_i \in \{0,1\} \text{ for all } i \in Q(\nu) \setminus T$$

Note that also for $i \in T$, the $x_i's$ are still to be decided. Applying the same procedure described in Section 3.2, it is easy to see that the optimal dual value for a generic subproblem is given by the solution of:

$$\min \sum_{i \in Q(\nu) \setminus T} (c_i x_i + x_i f_i(x_i)) + \sum_{i \in T} (c_i + x_i f_i(x_i)) \tag{42}$$
$$\text{s.t.} \quad \sum_{i \in Q(\nu)} x_i = 1 \tag{43}$$
$$x_i \geq 0 \text{ for all } i \in Q(\nu) \tag{44}$$

This problem can be efficiently solved using the Ordering Algorithm: just set $c_i = 0$ for all $i \in T$, because they must be in the solution since we pay their activation cost: the deeper is a node in the enumeration tree, the smaller is the problem to solve at that node, since the resources which are fixed OFF vanish from the problem.

The upper bound at the root node is given by the heuristic method described in Section 4.

### 5.2 Branching strategy

Two resources $u \in Q$ and $v \in Q$ are said *identical copies* if $c_u = c_v$ and $f_u(x) = f_v(x)$, $\forall\, x \in [0,1]$ (that resource is also said repeated). The number of copies of each resource (which is at least 1), can be found by suitably preprocessing the instance.

Branching here is to decide wether a resource $i$ is used and in how many copies. Clearly, at node $\nu$ of the enumeration tree, we branch on resources $i \in Q(\nu)\, T$. A possible branching rule—which turned out to be very effective— is to take $i \in Q(\nu) \setminus T$ such that $i = \arg\max\{c_i\}$.

We adopt an $n$-ary branching scheme. If the variable to be branched corresponds to a resource $i$ in $n$ copies, we generate $n+1$ nodes (subproblems): in the $l$-th subproblem, $l = 0, 1, \ldots, n$, $l$ copies of resource $i$ are ON and $n-l$ are off. Note that, with the $n$-ary branching, compared to a binary branching strategy (that is, not considering the impact of repeated resources on the solution), we only generate $n+1$ subproblems instead of $2^n$.

## 6 Computational experiments

The tests have been run on a PIV 3.2 GHz 1GB of RAM under Windows XP; our algorithms have been coded in C++. The computational experiments have been performed on the case where $f_i(x_i) = b_i x_i$, for this case being the minimal NP-hard case and to be able to compare it with CPLEX MIPQ solver.

## 6.1 Design of instances

We say that an instance is *non-dominated* if no resource is strictly preferable to any other, that is no $i, j$ exists such that $c_i + b_i < c_j + b_j$ and $c_i \leq c_j$. To this purpose we impose the following relations:

$$c_1 \geq c_2 \geq \cdots \geq c_q \qquad (45)$$
$$b_1 \leq b_2 \leq \cdots \leq b_q \qquad (46)$$
$$c_i + b_i > c_1 \quad \text{for all } i \in Q. \qquad (47)$$

We considered two classes of instances. The first is referred to as *base class* and it has the following properties: (i) ($b_i \neq b_j$) and ($c_i \neq c_j$), for all $1 \leq i < j \leq q$; (ii) $b_{i+1} = b_i + 1$ for all $1 \leq i < q$ and (iii) $b_i = f_{q-i+1}$, for all $1 \leq i \leq q$. Instances in this class are deterministically generated and there are no repeated resources. In the second class (the *random class*), instances are randomly generated and repeated resources are allowed. We imposed a computation time limit of 2 hours. Base class and random class instances are denoted, respectively, with a "b" and a "r" in the instances' name. We considered up to 1300 binary variables for the base class and up to 100 binary variables for the random class (due to the time limit constraint): for each number of variables in this class we generated 20 random instances.

## 6.2 Results and analysis

Experiments results are summarized in Table 1. Our algorithm outperforms CPLEX over all the instances classes and dimensions, except for very small problems (where the CPU time is very small indeed).

| instance | Our algo | | | CPLEX 10.0 | | |
|---|---|---|---|---|---|---|
| | CPU time | nodes # | iter. # | CPU time | nodes # | iter. # |
| b200 | 2.75 sec. | 2229 | 1114 | 1.97 | 4608 | 8112 |
| b400 | 17.13 sec | 10897 | 5448 | 27.55 | 42686 | 69655 |
| b600 | 66.89 sec | 34749 | 17374 | 184.95 sec. | 192012 | 304124 |
| b1000 | 515.45 sec | 192591 | 96295 | 2492.58 sec. | 1486308 | 2290007 |
| b1100 | 793 sec. | 274897 | 137493 | 4898.39 sec. | 2424233 | 3700426 |
| b1300 | 1774 sec. | 529275 | 264636 | >7200 sec. | >4000000 | >7000000 |
| r25 | 0.120 sec. | 106.2 | 36.2 | 0.060 sec. | 352.8 | 489.6 |
| r50 | 0.450 sec. | 800.8 | 134.6 | *106.370 sec. | *731558.5 | *874274.3 |
| r75 | 0.76 sec. | 1015.91 | 131.4 | *227.27 sec. | *1668124.9 | *1785904.7 |
| r100 | 5.3 sec. | 7774.45 | 633.1 | *821.35 sec. | *4575207.7 | *5277239.6 |

Table 1: Random instances (r) and base class instances (b); an asterisk means that CPLEX was stopped due to memory or CPU time limit (2 hrs). Data for random instances are average.

We first highlight the excellent performance of the heuristic procedure: in *all* the instances it found the optimal solution in a considerably small CPU time (always within seconds). As it can be seen from the dimension of the instances we tested, the more difficult ones are the random instances; in many cases CPLEX was stopped because of time limit or because of memory usage limit. This never happened for our algorithm. (The average values in those cases is computed considering the limit time or the time at that CPLEX stopped.) The reason why random instances are more critical is that repeated resources are allowed. Our multidimensional branching strategy turned out to be very effective, while CPLEX does not have any mechanism to avoid considering nodes which are exactly the replication of other nodes (when branching on repeated nodes). Figure 1 highlights the performance of our algorithm versus CPLEX: the bigger the instances the better is our algorithm.

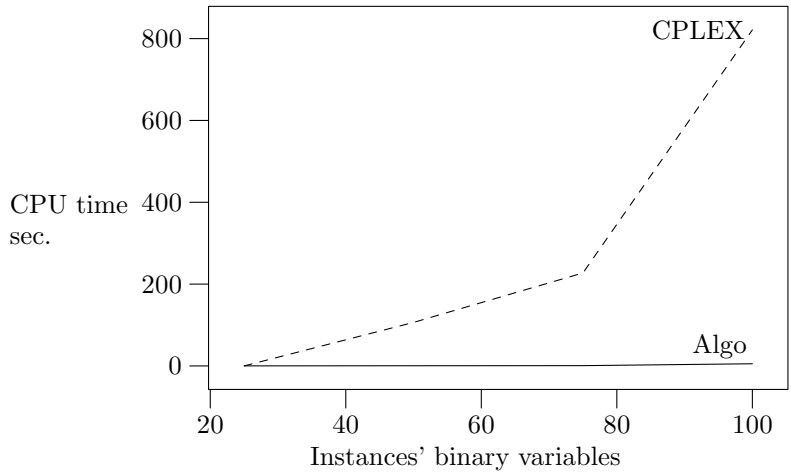

Figure 1: CPU time comparison for random instances. Our algorithm outperforms CPLEX.

# 7 Conclusions

In this paper we studied the problem of allocating a demand when the resources' costs are modeled with general convex functions. We devised a general solution approach based on a branch-and-bound scheme. The upper bound at the root node is given by a heuristic method whose perfomance is excellent (it found the optimum in all the tested instances). The lower bound at the leafs is a dual bound that is computed solving the continuous relaxation of the problem (their optimal values are identical) and for that task we devised an effective exact algorithm (the Ordering Algorithm). Computational experiments show that we outperform CPLEX and that our particular multidimensional branching strategy is very incisive.